\documentclass[12pt]{iopart}

\usepackage{iopams}
\usepackage{graphicx}
\usepackage{amsmath}
\usepackage{color}
\usepackage{xcolor}

\usepackage{textcomp}
\usepackage{lmodern}

\begin{document}

\title[Infrared laser magnetometry  with a NV doped diamond intracavity etalon]{Infrared laser magnetometry with a NV doped diamond intracavity etalon}

\author{Yannick Dumeige$^1$,  Jean-Fran\c cois Roch$^2$, Fabien Bretenaker$^{2,3}$, Thierry Debuisschert$^4$, Victor Acosta$^5$,  Christoph Becher$^6$, Georgios Chatzidrosos$^7$, Arne Wickenbrock$^7$, Lykourgos Bougas$^7$, Alexander Wilzewski$^7$, and Dmitry Budker$^{7,8,9}$}

 \address{$^1$ Univ Rennes, CNRS, FOTON - UMR 6082, F-22305 Lannion, France}
 \address{$^2$ Laboratoire Aim\'{e} Cotton, CNRS, Univ. Paris-Sud, ENS Cachan, Universit\'{e} Paris-Saclay, 91405 Orsay, France}
 \address{$^3$ Light and Matter Physics Group, Raman Research Institute, Bangalore 560080, India}
 \address{$^4$ Thales Research \& Technology, 1 avenue Augustin Fresnel, Palaiseau, France}
 \address{$^5$ University of New Mexico, Albuquerque, USA}
 \address{$^6$ Universit\"{a}t des Saarlandes, Germany}
 \address{$^7$ Johannes Gutenberg-Universit\"{a}t Mainz, 55128 Mainz, Germany}
 \address{$^8$ Department of Physics, University of California, Berkeley, CA 94720-7300, USA}
 \address{$^9$ Nuclear Science Division, Lawrence Berkeley National Laboratory, Berkeley, CA 94720, USA}
 \ead{yannick.dumeige@univ-rennes1.fr}
\vspace{10pt}
\begin{indented}
\item[]\today
\end{indented}

\begin{abstract}
We propose an hybrid laser system consisting of a semiconductor external cavity laser associated to an intra-cavity diamond  etalon  doped with nitrogen-vacancy color centers. We consider laser emission tuned to the infrared absorption line that is enhanced under the magnetic field dependent nitrogen-vacancy electron spin resonance and show that this architecture leads to a compact solid-state magnetometer that can be operated at room-temperature. The sensitivity to the magnetic field limited by the photon shot-noise of the output laser beam is estimated to be around  $250~\mathrm{fT/\sqrt{Hz}}$. Unlike usual NV center infrared magnetometry, this method would not require an external frequency stabilized laser. Since the proposed system relies on the competition between the laser threshold and an intracavity absorption, such laser-based optical sensor could be easily adapted  to a broad variety of physical systems.
\end{abstract}

%\pacs{07.55.Ge, 76.30.Mi, 42.55.Px}
\vspace{2pc}
\noindent{\it Keywords}: diamond NV center,  optical magnetometry, VCSEL
%\submitto{\JPD}
%\maketitle

\section{Introduction}

%Based on the optically detected magnetic resonance (ODMR) of its electron spin, the  nitrogen-vacancy (NV) center  in diamond is a solid-state atomic magnetometer.
In recent years, the optical detection of the magnetic resonance between the electronic triplet $S=1$ spin states of the negatively charged nitrogen-vacancy (NV) color center in diamond and the measurements of the Zeeman shifts induced by an applied magnetic field  has been used in a variety of solid-state magnetometers \cite{Rondin14}. Due to the special properties of the NV center, these systems can be operated in ambient conditions to detect a broad range of magnetic fields created by both  physical and biological systems  \cite{Schirhagl14,CasolaYacobi18}. By raster scanning a single  NV spin over a magnetized substrate and detecting the spin-dependent luminescence emitted by this atomic-like defect, the stray magnetic  field created by the sample magnetization can be mapped with nanometer spatial resolution \cite{Balasubramanian08,Maze08}. Compared to a single spin, the  magnetic field sensitivity of an ensemble of NV centers contained in a macroscopic single-crystal diamond sample is increased by $\sqrt{N}$ where $N$ is the number of NV centers used as magnetic sensors \cite{Acosta09}. Due to this enhancement, continuous-wave magnetometry based on an NV ensemble has recently reached a sensitivity  level of about  ${15~\rm pT}/{\sqrt{\rm Hz}}$ \cite{Barry16, Schloss18}. However this technique is constrained by the  collection efficiency of the NV luminescence  and by parasitic background  light that spectrally overlaps the  broad luminescence of the NV center with wavelength extending from  637 nm  to about 800 nm. 

The spin state of the NV center can also be determined by  the infrared (IR) optical transition associated with the  singlet $S=0$ spin state \cite{Rogers08,Acosta10b, Kehayias13}. The corresponding scheme for detecting the perturbation by an applied magnetic field is based  on the absorption of  a signal beam   tuned to this IR transition centered at $\lambda_s= 1042\,{\rm nm}$ wavelength \cite{Acosta10}. The magnetic-field dependent  signal  is then free from any background  and the relevant photon detection efficiency can be almost ideal. Nevertheless, the low optical depth of the IR transition at room temperature, even for a dense ensemble of NV centers,  needs to be compensated by a multi-pass configuration \cite{Clevenson15}. This enhancement scheme  can be  implemented  by placing the NV doped diamond in an  optical cavity resonant with the IR signal beam \cite{Dumeige13,Jensen14}. A shot-noise limited sensitivity of $28~\mathrm{pT/\sqrt{Hz}}$ was recently achieved  using a miniaturized Fabry-Perot cavity \cite{Chatzidrosos17}  which could even  be  realized in integrated diamond photonics \cite{Gazzano17,Bougas18}.

In order to circumvent the previously mentioned drawbacks of luminescence based magnetometers, it was proposed   to  operate the NV center  transition between the triplet spin states in the stimulated emission regime  \cite{Faraon12,Jeske16,Savitski17} so that population inversion in the NV center levels  provides the optical gain of a laser. By setting the laser at its threshold,  sensitivities of about  $\mathrm{fT/\sqrt{Hz}}$ are anticipated \cite{Jeske16}. 
Nevertheless the stimulated emission from the NV  centers can be strongly affected by the excited state absorption (ESA) phenomena and  by the photoconversion  between the negatively-charged state   $\mathrm{NV}^{-}$, with  the previously described spin triplet structure, and the  neutral charge state $\mathrm{NV}^{0}$ \cite{Subedi18}. These parasitic effects can make  the implementation of NV$^-$ center  magnetometry based on the visible optical laser amplification challenging 
\cite{Jeske17}.
  
Here we propose to combine the IR absorption method and the laser threshold magnetometry method by considering  a hybrid laser architecture which integrates the  diamond sample containing the NV centers in an external-cavity laser. The optical gain in the laser is provided by an independent semiconductor material which is optically pumped. The laser threshold of the whole system is then sensitive to the applied magnetic field via the  losses on the IR transition induced by the spin resonance of the NV centers. In this scheme, the  ESA in the gain medium becomes irrelevant and has a marginally negative effect on the IR signal absorption efficiency. 
 %This architecture also avoids to use an additional IR frequency stabilized laser as  it is required for the  cavity enhancement schemes \cite{Dumeige13,Jensen14,Chatzidrosos17}. 
Using a rate equation model of the photodynamics of the NV center that takes into account its two charge states, we  evaluate  the magnetic field sensitivity of this hybrid laser system. Finally, we discuss the possible advantages of this sensor architecture for practical applications.

\section{Model of the spin-dependent NV center dynamics}

\begin{figure}[htbp]
\centering\includegraphics[width=12cm]{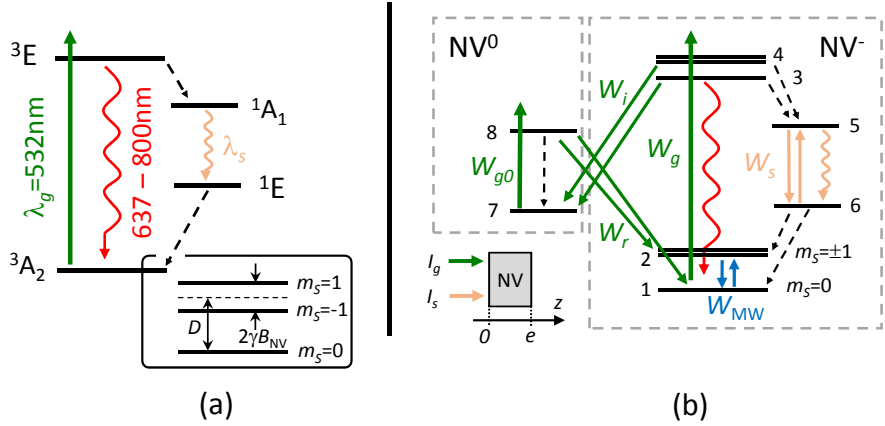}
\caption{Energy diagram of the NV center. (a) Level scheme of the NV$^-$ center. As shown in the insert, the ground state $^3A_2$ is split into $m_S=0$ and $m_S=\pm 1$ sublevels due to spin-spin interaction and 
an external magnetic field applied to the NV center lifts the $m_S=\pm 1$ degeneracy. The NV$^-$ center is spin polarized into $m_S=0$ by optical pumping at $\lambda_g=532~\mathrm{nm}$. The resonance zero-phonon wavelength of the singlet transition is $\lambda_s=1042~\mathrm{nm}$. Typical lifetimes of the $^3E$ and $^1E$ levels are respectively $16~\mathrm{ns}$ and $600~\mathrm{ns}$. (b) Description of the  photodynamics between the spin sublevels of the NV$^-$ and NV$^0$ ground and excited electronic states. $W_g$ is the pumping rate associated to the $\mathrm{NV}^{-}$ and $W_{g0}$ that of the $\mathrm{NV}^{0}$. $W_s$ is the transition rate of the IR resonance. $W_i$ and $W_r$ are respectively the ionization and recombination rates of the $\mathrm{NV}^{-} \rightleftarrows \mathrm{NV}^{0}$ transition. $W_{\rm MW}$ is the $m_S=0 \rightleftarrows m_S=\pm1$ transition rate induced by the resonant microwave field.  The insert shows the pump (green) and signal (IR) configuration, with propagation  through  the diamond plate of thickness $e$.}
\label{fig_energy}
\end{figure}

The NV  center consists of a nitrogen impurity linked to an adjacent vacant lattice site. In the negatively charged state NV$^-$ which consists of six electrons associated to the dangling bonds around the lattice vacancy, four of these electrons populate the lowest energy states \cite{Doherty13}. The remaining two electrons create both spin triplet $S=1$ states and spin singlet $S=0$ states that are associated to optical transitions within the $\rm 5.5~eV$ bandgap of diamond. In the spin triplet manifold of the ground electronic state $^{3}A_2$, the magnetic interaction between electron spins induces a zero-field splitting of $D\approx  2.87\,{\rm GHz}$ between the $m_S=0$ and $m_S=\pm 1$  spin projection sublevels along the  intrinsic quantization axis that is defined by the N-to-\!V axis of the defect inside the crystal (Fig. \ref{fig_energy}a). 

According to selection rules determined from group-theory methods  \cite{Doherty13,Maze11}, the optically electronic transitions between the triplet sublevels of the  $^{3}A_2$  electronic ground state and the corresponding  triplet sublevels of  the  excited electronic level $^{3}E$ are  mainly  spin-conserving.
 %\cite{Doherty13,Maze2011}.  
Due to  a non-radiative decay path  from  the  $m_S=\pm 1$  excited  states  through  the metastable  singlet states $^{1}A$ and $^{1}E$ and then preferentially back to the $m_S=0$ ground state (Fig. \ref{fig_energy}a), green laser optical excitation  of the $^{3}A_2$ triplet sublevels   polarizes the electron spin of the NV$^-$ center  into the $m_S=0$ sublevel \cite{Thiering18}. The non-radiative leakage to the metastable $S=0$ state also induces  a lower luminescence efficiency of the $m_S=\pm 1$ sublevels compared to  $m_S=0$ so that the  occupation probability in this ground state spin manifold  of $m_S=\pm 1$ compared that of the $m_S=0$   can be determined by  monitoring the  photoluminescence (PL) intensity. These properties enable the optically detected magnetic resonance (ODMR) signal that can be induced by applying a  microwave field resonant with the $m_S=0$ to $m_S=\pm 1$ transition. Since a magnetic field applied to the NV center induces Zeeman  shifts  that lift the degeneracy of the $m_S=\pm 1$ sublevels, the magnetic field amplitude can be determined by measuring these Zeeman shifts in the ODMR microwave frequency spectrum  \cite{Taylor08}. 

The detection of spin polarization can also be realized by measuring the transmission of a signal IR beam that probes the absorption on the  transition between the singlet metastable states $^{1}A$ and $^{1}E$  \cite{Acosta10}. Under green light continuous optical pumping and in the absence of resonant microwaves, the NV centers are   pumped   into   the $m_S=0$ ground sublevel leading to a reduced occupation rate of the metastable singlet state $^{1}E$. In this off-resonance regime the IR signal transmission is maximal. For magnetic fields applied along one of the NV axis, when the microwave field frequency is resonant   with   frequency $ D \pm \gamma B_{\rm NV}/(2\pi)$, 
 where $ \gamma=1.761 \times 10^{11} \; {\rm rad}\; {\rm s}^{-1} \; {\rm T}^{-1}$ is the NV gyromagnetic ratio and $B_{\rm NV}$ the projection on the NV axis of the applied magnetic field,  the population is transferred into the $m_S = \pm 1$ ground state. The   occupation rate of the   $^{1}E$ state increases and the  magnetic field dependent spin resonance  can be detected as a lower  transmission of the  IR signal beam.

A rate equation model is used to describe the photodynamics of the triplet and singlet states and to estimate the optical losses induced by the magnetic resonance between the sublevels of the $^{3}A_2$ ground state on the signal beam that propagates through the NV doped diamond sample \cite{Dumeige13, Bougas18}. In order to take into account the photoionization process \cite{Dumeige04, Manson06, Aslam13, Meirzada17} between  $\mathrm{NV}^{-}$ and $\mathrm{NV}^{0}$ two supplementary levels associated to the $\mathrm{NV}^{0}$ neutral charge state \cite{Meirzada17} are added to this level scheme, as  shown in Fig. \ref{fig_energy}b. In our configuration, the photoionization and the ESA only reduce the numerical value of the inferred IR absorption cross-section (see \ref{donnee}) but are not an intrinsic limitation as it is the case for laser threshold magnetometry based on the visible transition.

The spin sublevels $m_S=0$ and $m_S=\pm 1$ of the $^3 A_2$  state of the $\mathrm{NV}^{-}$ center  are labelled $1$ and $2$ whereas $3$ and $4$ are the corresponding spin sublevels of the excited state $^3 E$. 
The  ground and excited states of the singlet IR transition are respectively labelled $6$ and $5$. Finally, $7$ and $8$ are the $\mathrm{NV}^{0}$ ground and excited states. The radiative or non-radiative relaxation rate from $\alpha$ to $\beta$ levels is $k_{\alpha\beta}$; the values of these parameters are given in \ref{donnee} and similar measurements can be found in \cite{Robledo11, Kalb18}. The relaxation rate from $2$ to $1$ can be neglected since the associated spin-relaxation time, longer than $0.2~\mathrm{ms}$ at room temperature \cite{Mrozek15}, is much longer than  all other decay processes. When excited in the upper singlet state, the system can only decay to the lower singlet state and thus $k_{51}=k_{52}=0$. Finally, the optical transition are spin conserving and thus $k_{41}=k_{32}=0$. The optical depth of a diamond plate of thickness $e$ doped with the NV centers (see the insert of Fig. \ref{fig_energy}b) is obtained from the steady state solution of the following system calculated at each position  indexed by $z$ in the diamond sample:    
%As $k_{35}\ll k_{45}$, the $\mathrm{NV}^{-}$ is spin polarized in $m=0$ by optical pumping. This induces a reduction of the population of level $6$ leading to a low IR absorption. The spin of the ground state can be driven to $m=\pm 1$ using a resonant microwave field with frequency $D\pm\gamma B/(2\pi)$ (where $B$ is the projection of the magnetic field along one of the four $\mathrm{NV}^{-}$ orientations and $\gamma=1.761\times 10^{11}~\mathrm{rad~s^{-1}~T^{-1}}$ the gyromagnetic ratio) which gives a significantly increase of the IR light absorption by level $6$. 
\begin{equation}
  \left\{
    \begin{aligned}
      \frac{dN_1}{dt} & =-(W_g+W_{\rm MW})N_1+W_{\rm MW}N_2+k_{31}N_3+k_{61}N_6+\frac{W_r}{2}N_8\\
      \frac{dN_2}{dt} & =W_{\rm MW}N_1-(W_g+W_{\rm MW})N_2+k_{42}N_4+k_{62}N_6+\frac{W_r}{2}N_8\\
      \frac{dN_3}{dt} & =W_g N_1-(k_{31}+k_{35}+W_i)N_3\\
      \frac{dN_4}{dt} & =W_gN_2-(k_{42}+k_{45}+W_i)N_4\\
			\frac{dN_5}{dt} & =k_{35}N_3+k_{45}N_4-(k_{56}+W_s)N_5+W_sN_6\\
			\frac{dN_6}{dt} & =(k_{56}+W_s)N_5-(W_s+k_{61}+k_{62})N_6\\
			\frac{dN_7}{dt} & =W_iN_3+W_iN_4-W_{g0}N_7+k_{87}N_8\\
			\frac{dN_8}{dt} & =W_{g0}N_7-(k_{87}+W_r)N_8,
    \end{aligned}
  \right.
\end{equation}
where $N_{\alpha}(z)$ is the population density of state $\alpha$. The pumping rates are related to the pump $I_g$ and signal $I_s$  optical intensities through $W_g=\sigma_gI_g\lambda_g/(hc)$, $W_{g0}=\sigma_{g0}I_g\lambda_g/(hc)$, $W_i=\sigma_iI_g\lambda_g/(hc)$, $W_r=\sigma_rI_g\lambda_g/(hc)$ and $W_s=\sigma_sI_s\lambda_s/(hc)$, where  the cross-sections $\sigma_{\beta}$ are given in \ref{donnee}. The system is considered as closed and $\sum_{\alpha=1}^8 N_{\alpha}(z)=N_{\rm NV}$ where $N_{\rm NV}$ is the  density of the NV centers contained in the diamond sample. The pump (green) and signal (IR) intensities at the output of the diamond sample are then obtained by:  
\begin{equation}
  \left\{
    \begin{aligned}
      \frac{dI_g}{dz} & =-\left[\sigma_g(N_1+N_2)+\sigma_{g0}N_7+\sigma_i(N_3+N_4)+\sigma_rN_8\right]I_g\\
      \frac{dI_s}{dz} & =-\sigma_s(N_6-N_5)I_s.\\
     \end{aligned}
  \right.
\end{equation}
After integration along $z$, these equations determine  the optical depth $\tau$ for the IR signal beam as a function of the  green-light $W_g$ and microwave $W_{\rm MW}$ pumping rates:
\begin{equation}
\tau=-\ln{\left[\frac{I_s(e)}{I_s(0)}\right]}.
\end{equation}
 
\section{Hybrid architecture for NV laser magnetometry}

\begin{figure}[htbp]
\centering\includegraphics[width=12cm]{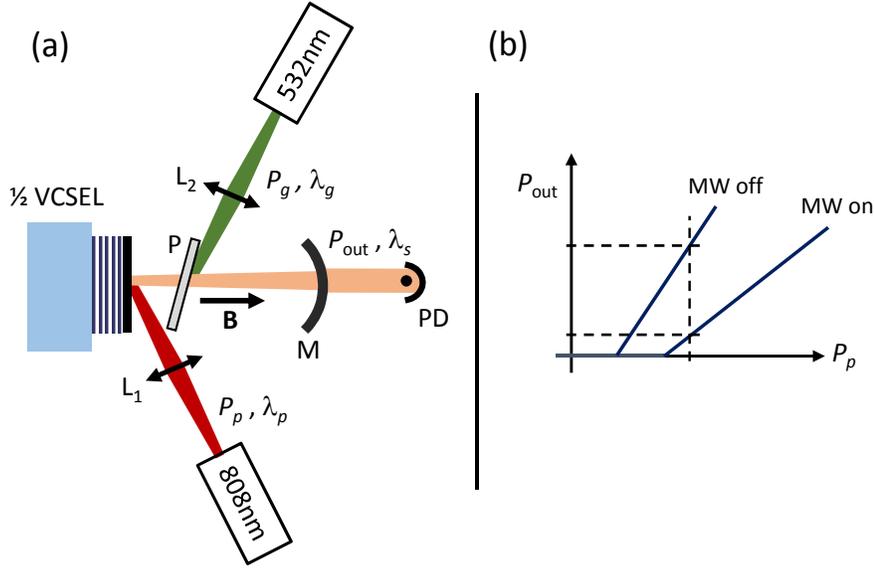}
\caption{(a) Hybrid magnetometer architecture combining a half-VCSEL, a diamond thin plate (P) highly doped with NV centers, and an output coupling mirror (M). $\mathrm{L_1}$ and $\mathrm{L_2}$ are two focusing lenses. P is the diamond plate containing the NV centers. $P_g$, $P_p$ and $P_{\rm out}$ are respectively the power for NV polarization, the power for quantum well pumping, and the output  power of the IR laser  which is detected by a photodiode PD. The half-VCSEL  represents the Bragg mirror and the semiconductor quantum wells which provide the optical gain in the laser cavity. (b) Operation principle of the magnetometer showing the threshold and the efficiency of the external cavity laser with the microwave (MW) field being either on-resonance or off-resonance.}
\label{fig_principle}
\end{figure}

The proposed hybrid architecture shown in Fig. \ref{fig_principle}a is based on a vertical external cavity surface emitting laser (VECSEL). The gain medium is a half- vertical cavity surface emitting laser (VCSEL) consisting of semiconductor multiple InGaAs/GaAs quantum wells grown on a perfectly reflecting Bragg mirror both centered at $\lambda_s$ \cite{Baili07}. The output coupling mirror (M) of the laser cavity has a transmission coefficient $T$. A diamond thin plate containing a high concentration of NV centers is inserted inside the cavity. The semiconductor quantum wells are pumped using a laser at $\lambda_p=808~\mathrm{nm}$ and the NV centers are spin polarized by illuminating the diamond sample with a green laser at $\lambda_g=532\, {\rm nm}$ wavelength. The diamond plate operates as an intracavity etalon leading to single-mode operation of this external cavity semiconductor laser. The extra losses due to the NV absorption in the diamond plate, and thus the threshold and the efficiency of this hybrid laser depend on the spin state of the NV centers that are driven by the microwave field. Consequently, the output power $P_{\rm out}$ of the laser  
 can be modified by the  magnetic field $\mathbf{B}$   applied on the NV centers. As previously explained, the IR losses are increased when the microwave field is on-resonance, leading to a higher threshold and a lower efficiency compared to the off-resonance case as shown in Fig. \ref{fig_principle}b. Using these   features  the magnetic field dependent spin resonance can  be detected by monitoring the IR laser output power.

\section{Parameters of the VECSEL}\label{VECSEL}

The main requirements on the IR laser are (i) to operate in the regime of high-finesse cavity in order to increase the effective path of the IR signal in the diamond plate \cite{Dumeige13, Jensen14, Chatzidrosos17}, (ii) to be low-noise since the magnetic field sensitivity is directly related to the IR optical signal noise and (iii) compactness. The VECSEL-based architecture is therefore  a good candidate especially when the cavity length is chosen to reach the class A regime of laser operation (corresponding to a cavity lifetime longer than population inversion lifetime) enabling a photon shot-noise limited amplitude noise operation\cite{Baili09}.

The parameters of the hybrid laser magnetometer are deduced  from those given in Ref. \cite{Baili07} which describes a shot-noise limited semiconductor VECSEL emitting at a wavelength of $1~\mu\mathrm{m}$, close to $\lambda_s$. In the class A regime, the output power $P_{\rm out}$ of the IR laser is given by:
\begin{equation}\label{Output_power}
P_{\rm out}=TP_{\rm sat}(r-1),
\end{equation}
where $P_{\rm sat}$ is the pumping saturation power, and $r$ the  rate of the pumping power $P_p$ above the laser threshold  $P_{\rm th}$: 
\begin{equation}\label{seuil1} 
r=\frac{P_p}{P_{\rm th}} = \frac{\eta P_p}{T+\epsilon} \,, 
\end{equation}
where $\epsilon$ are the losses introduced by the intracavity etalon  for  a round trip  inside the cavity, and   $\eta$ is the proportionality factor that relates the optical gain obtained after one  round trip in the cavity to the pumping power $P_p$. 
 
With an intracavity etalon that ensures single-mode operation, the laser realized in Ref.\,\cite{Baili07} has a threshold power of $P_{\rm th}=700~\mathrm{mW}$ and provides an output power of $P_{\rm out}=50~\mathrm{mW}$ for $P_p=1~\mathrm{W}$ of  pump power applied to the VECSEL. Considering an   output coupling mirror with transmission $T=1~\%$, we then infer from Eq. (\ref{Output_power})  a saturation power of $P_{\rm sat}=11.7\,\mathrm{W}$. Without the intracavity
 etalon, the output power is $P_{\rm out}=140\,\mathrm{mW}$ for the same  pump power. Since in this case $\epsilon=0$, we  deduce $\eta=2.2\times 10^{-2}\,\mathrm{W}^{-1}$ by combining Eq. (\ref{Output_power}) and Eq. (\ref{seuil1}). If we   consider again the case of the etalon in the laser cavity, we have at the threshold $\eta P_{\rm th}=T+\epsilon$ so that  $\epsilon=0.5~\%$.  

\section{Intracavity diamond  etalon and  magnetic field sensitivity}

We now consider that the intracavity etalon consists of a diamond sample doped with NV centers, without any anti-reflection coating on the input and output facets. The etalon is illuminated using an additional green laser which  polarizes the NV spins  in the $m_S=0$ sublevel of the ground electronic state and also feeds the metastable singlet level (6) shown in Fig. \ref{fig_energy}b. Taking into account the optical thickness $\tau\ll 1$ of the diamond plate, the absorption of the IR beam due to the singlet transition of the NV centers then   corresponds to   additional intracavity optical losses
\begin{equation} \label{loss_diamond}
\xi=2\chi\tau
\end{equation}
where  the factor 2 accounts for the round trip inside the cavity and  $\chi=(n_d^2+1)/(2n_d)\approx 1.4$ is an enhancement factor of the losses which is induced by   the  high refractive  index  $n_d=2.4$ of the diamond plate (see \ref{enhancement}). The laser pumping rate  then becomes:
\begin{equation}\label{new_r}
r=\frac{\eta P_p}{T+\epsilon+\xi} 
\end{equation}
where  $\eta$ and $\epsilon$ have the values previously determined. 
The parameter $\xi$ corresponds to the useful losses of the diamond sample that  determine the efficiency of the laser response to the applied magnetic field, as:
\begin{equation}
 \frac{\partial P_{\rm out}}{\partial B} =\frac{\partial P_{\rm out}}{\partial \tau}\cdot \frac{\partial \tau}{\partial \nu_{\rm ESR}}\cdot \frac{\partial \nu_{\rm ESR}}{\partial B}
\end{equation}
where $\nu_{\rm ESR}$ is the resonance frequency of the microwave field with the dependence $\frac{\partial \nu_{\rm ESR}}{\partial B}= \frac{\gamma}{2\pi}$ to the applied magnetic field. If we assume that the spin resonance has a Lorentzian lineshape, the maximum of $\frac{\partial \tau}{\partial \nu_{\rm ESR}}$ is reached for $\tau_{\rm max}=(\tau_{\rm on}+3\tau_{\rm off})/4$  where $\tau_{\rm on}$ and $\tau_{\rm off}$ are the optical depths with respectively the microwave field being either on-resonance or off-resonance (see  \ref{Odepth}). This maximum value is then  given by
\begin{equation}\label{max_sens}
\left|\frac{\partial \tau}{\partial \nu_{\rm ESR}}\right |_{\rm max}=\frac{3\sqrt{3}}{4} \, \frac{\Delta\tau}{\Delta \nu_{\rm ESR}},
\end{equation}
where $\Delta\tau=\left|\tau_{\rm off}-\tau_{\rm on}\right|$ and $\Delta \nu_{\rm ESR}$ is the full width at half maximum of the spin resonance. We assume here that the linewidth of the electronic spin resonance (ESR) is limited by the spin dephasing time $T_2^*$ and by the spin polarization relaxation rate $\Gamma$ taking into account populations dynamics \cite{Dreau11} and related to $W_{\rm sat}$ the microwave saturation rate by $\Gamma=2W_{\rm sat}$, we thus can write
\begin{equation}\label{nuESR}
\Delta \nu_{\rm ESR}=\frac{1}{\pi T_2^*}\sqrt{1+\frac{\Omega_R^2T^*_2}{\Gamma}},
\end{equation}
where $\Omega_R$ the Rabi frequency is related to the microwave pumping rate by $W_{\rm MW}=\frac{\Omega_R^2T^*_2}{2}$. Taking into account the pumping rate given by Eq. (\ref{new_r}), we can then determine the maximal response  of the laser-based magnetometer:
\begin{equation}\label{sensibilite}
\left|\frac{\partial P_{\rm out}}{\partial B}\right|_{\rm max}=\frac{3\sqrt{3}}{2} \chi\, \Delta\tau\,  \, \frac{\gamma}{2\pi\Delta \nu_{\rm ESR}}\, T \, P_{\rm sat} \, \frac{\eta P_p}{(T+\epsilon+\xi_{\rm max})^2} \,
\end{equation}
with $\xi_{\rm max} = 2 \chi \tau_{\rm max}$. Assuming that the laser output noise is at the limit of photon shot-noise we have
\begin{equation}\label{SN}
\delta P_{\rm out}=\sqrt{\frac{P_{\rm out}hc \Delta f}{\lambda_s}},
\end{equation}
where $\Delta f$ is the measurement bandwidth. The equivalent magnetic noise of the sensor $\delta B_{\rm min}=\frac{\delta P_{\rm out}}{\left|\partial P_{\rm out}/\partial B\right|_{\rm max}}$ can then be deduced from Eqs. (\ref{nuESR}), (\ref{sensibilite}) and (\ref{SN}).
%The magnetic field noise  of the sensor limited by the output power of the laser is given by
%\begin{equation}\label{def}
%\delta B=\frac{\delta P_{out}}{\left|\frac{\partial P_{out}}{\partial B}\right|}.
%\end{equation}

\section{Results}\label{results}

The simulations of the equivalent magnetic field noise are based on the laser parameters given in section \ref{VECSEL} apart from  the diamond etalon  with  thickness $e=100~\mu\mathrm{m}$ (note that we take into account parasitic losses due to diamond by taking $\epsilon=0.5~\%$). We also assume that the laser emission is tuned to the NV center IR transition.
\begin{figure}[htbp]
\centering\includegraphics[width=13.5cm]{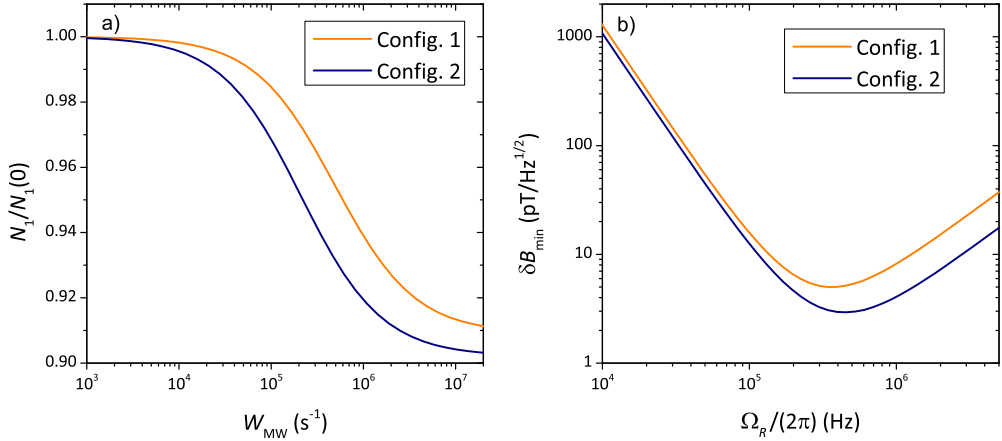}
\caption{a) Normalized population $N_1$ versus microwave pumping rate $W_{\rm MW}$. b) Equivalent magnetic field noise for an IR NV center laser magnetometer versus Rabi frequency $\Omega_R$ of the microwave field. Config. 1: $N_{\rm NV}=4.4\times 10^{23}~\mathrm{m}^{-3}$ and  $T_2^*=390~\mathrm{ns}$. Config. 2:  $N_{\rm NV}=2.8\times 10^{24}~\mathrm{m}^{-3}$ and $T_2^*=150~\mathrm{ns}$. For both figures, calculations have been carried out for $I_g=40~{\rm kW}\, {\rm cm}^{-2}$. For the calculations of the equivalent magnetic field noise the on-resonance optical depth is calculated using the actual value of $\Omega_R$ whereas the off-resonance value is obtained for $\Omega_R=0$. Note that for the calculation of on-resonance optical depths, we consider that only 1/4  of the NV centers are aligned along the magnetic field. Furthermore in Fig. \ref{Rabi}b) we used the following laser parameters: $T=0.03$ and $r=1.2$ which corresponds to an unoptimized value of the laser pumping rate.}
\label{Rabi}
\end{figure}
We now consider two configurations with  different  realistic densities of NV centers. Config. 1 refers to $N_{\rm NV}=4.4\times 10^{23}~\mathrm{m}^{-3}$ and $T_2^*=390~\mathrm{ns}$ \cite{Kubo11} whereas Config. 2 to $N_{\rm NV}=2.8\times 10^{24}~\mathrm{m}^{-3}$ and $T_2^*=150~\mathrm{ns}$ \cite{Acosta09}. The length of the cavity and the curvature of the output mirror are such that the waist of the laser mode is $w_0=50~\mu\mathrm{m}$. The diamond etalon is located as close to the waist position as allowed by the pumping beam. We assume a green pump intensity $I_g=40~{\rm kW}\, {\rm cm}^{-2}$ corresponding to a mean power of $1.5~\mathrm{W}$. 

\begin{figure}[htbp]
\centering\includegraphics[width=15cm]{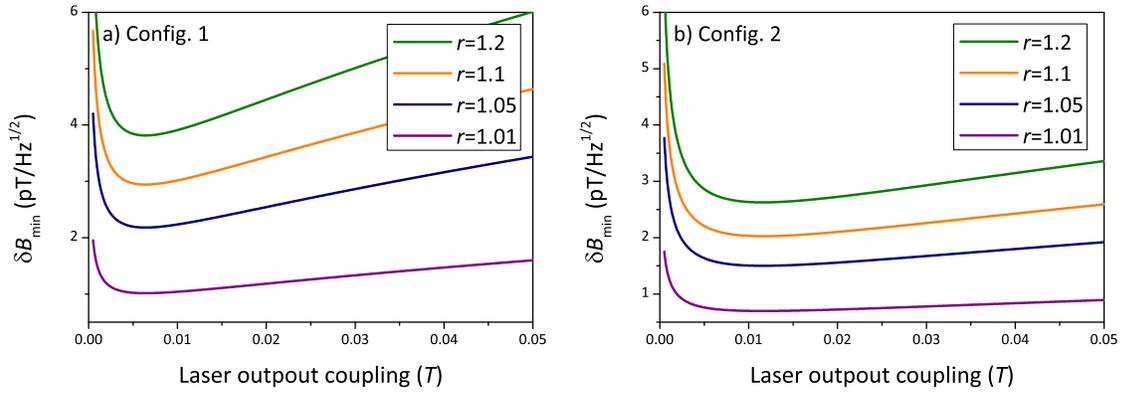}
\caption{Equivalent magnetic field noise optimization of the IR NV center laser magnetometer obtained for $I_g=40~\mathrm{kW/cm^2}$. (a) Config. 1, $\Omega_R=2\pi\times 3.4\times 10^5~\rm Hz$. (b) Config. 2, $\Omega_R=2\pi\times 4.5\times 10^5~\rm Hz$.}
\label{fig_plot_results}
\end{figure}

We first show in Fig. \ref{Rabi}a) the population $N_1$ as a function of the microwave pumping rate for the two configurations. By fitting the results by $A+\frac{B}{1+W_{\rm MW}/W_{\rm sat}}$ with $A$, $B$ and $W_{\rm sat}$ as free parameters we can deduce the microwave saturation rate, for Config. 1: $W_{\rm sat}=4.9\times 10^5~\rm s^{-1}$ and for Config. 2: $W_{\rm sat}=2.1\times 10^5~\rm s^{-1}$. We then are able to plot the equivalent magnetic field noise $\delta B_{\rm min}$ versus the microwave Rabi frequency for the two studied configurations. In both cases, the equivalent magnetic field noise reaches an optimum coming from the trade-off between the increase of the contrast and the broadening of the ESR. The following optimal Rabi frequencies are used in the rest of the work: $\Omega_R=2\pi\times 3.4\times 10^5~\rm Hz$ for Config. 1 and $\Omega_R=2\pi\times 4.5\times 10^5~\rm Hz$ for Config. 2. Further optimization of the results are shown in Fig. \ref{fig_plot_results} where the equivalent magnetic field noise $\delta B_{\rm min}$ is plotted as a function of the transmission of the output mirror for several pumping rates $r$. For both configurations an optimum output coupling is found depending on the IR absorption. Figure \ref{fig_plot_results} also shows  that by operating the laser close to its threshold (here $r=1.01$), the equivalent magnetic field noise can be strongly reduced, reaching for instance almost $700~\mathrm{fT}/\sqrt{\mathrm{Hz}}$ for the parameters of Config. 2. This value could be reduce to $250~\mathrm{fT}/\sqrt{\mathrm{Hz}}$ by using techniques to avoid ESR broadening due to microwave pumping \cite{Dreau11} and considering a higher Rabi frequency $\Omega_R=2\pi\times 1~\rm MHz$.
Indeed, at its threshold, the laser becomes highly sensitive to the intracavity optical losses and thus to magnetic field fluctuations similarly to the behavior of  visible laser threshold magnetometry  \cite{Jeske16}. Finally  comparison between the two configurations of Fig. \ref{fig_plot_results} shows that the trade-off between the NV center density and the spin dephasing time associated with Config. 2 leads to an improved sensitivity. Note that once fundamental limits are achieved, the sensitivity scales as $1/\sqrt{N_{\rm NV}T_2^*}$ \cite{Taylor08}. For the considered diamond thickness and waist size the spin projection noise determined by the total number of NV centers participating in the measurement is smaller than $30~\mathrm{fT}/\sqrt{\mathrm{Hz}}$ \cite{Taylor08}. This noise can therefore be neglected compared to the shot-noise limit set by the laser output photon flux. Note the spin projection noise limit could be reached by operating closer to the laser threshold.

\section{Conclusion}

We have shown that magnetometry based on the   IR absorption associated to the singlet states of the NV$^-$ center can be implemented by integrating a diamond sample containing  the NV centers inside an external half-VCSEL cavity. This scheme does not require a narrow linewidth stabilized IR laser as in  realizations  based on multi-pass absorption in a resonant passive cavity  \cite{Chatzidrosos17}. Compared to previous proposals consisting of a diamond laser using the NV$^-$ centers for optical amplification, the detrimental effects of both the parasitic ESA by the triplet excited state and the photoconversion to the NV$^0$ charge state are also circumvented since the optical gain is obtained from  an independent system. Moreover, the use of a semiconductor material makes it possible to consider electric-current pumping which is of great interest for practical implementations avoiding the pump/signal configuration \cite{Chatzidrosos17, Savitski17}. Our simulations show that a photon shot-noise limited sensitivity of about $700~\mathrm{fT}/\sqrt{\mathrm{Hz}}$ (and even $250~\mathrm{fT}/\sqrt{\mathrm{Hz}}$ if the ESR linewidth is limited by the spin dephasing time) can be reached for realistic parameters.

\section*{Acknowledgement}

We acknowledge Isabelle Sagnes for fruitful discussions on the VCSEL fabrication. The work of JFR, FB, and TD is performed in the framework of the joint research lab between
Laboratoire Aim\'{e} Cotton and Thales R\&T. This project has received funding from the  European Union Seventh Framework Programme (FP7/2007-2013) under the project DIADEMS (grant agreement No.611143), the German Federal Ministry of Education and Research (BMBF) within the Quantumtechnologien program (FKZ 13N14439) and from CNRS under the PICS project MOCASSIN. YD acknowledges the support of the Institut Universitaire de France and the Alexander von Humboldt Foundation.

\appendix

\section{Photophysical parameters}\label{donnee}

Table \ref{tableau} gives the values of the photophysical parameters used in the simulations. Since we considered the transition between the two charge states NV$^-$ and NV$^0$, we  updated the value of the IR absorption cross-section which was previously inferred from experimental data \cite{Dumeige13}. For this purpose, we used the same method consisting in adjusting the value of $\sigma_{s}$ to obtain the experimental value of the single-pass IR transmission reported in \cite{Acosta10}. 
\begin{table}[h!]
\caption{\label{photo_tab}Physical parameters (defined in Fig. \ref{fig_energy}) used to model the NV center optical depth at $\lambda_s$. The uncertainties on $\sigma_s$ is calculated to obtain an overlap with previous estimations \cite{Dumeige13}.} 
\begin{indented}\label{tableau}
\lineup
\item[]\begin{tabular}{@{}*{7}{l}}
\br                              
Parameter&Value&Reference\cr 
\mr
$k_{31}=k_{32}$&$(66\pm5)~\mu\mathrm{s}^{-1}$&\cite{Tetienne12}\cr
$k_{35}$&$(7.9\pm4.1)~\mu\mathrm{s}^{-1}$&\cite{Tetienne12}\cr
$k_{45}$&$(53\pm7)~\mu\mathrm{s}^{-1}$&\cite{Tetienne12}\cr
$k_{61}$&$(1.0\pm0.8)~\mu\mathrm{s}^{-1}$&\cite{Tetienne12}\cr
$k_{62}$&$(0.7\pm0.5)~\mu\mathrm{s}^{-1}$&\cite{Tetienne12}\cr
$k_{56}$&$1.0~\mathrm{ns}^{-1}$&\cite{Acosta10}\cr
$k_{87}$&$(53\pm7)~\mu\mathrm{s}^{-1}$&\cite{Meirzada17}\cr
$\sigma_{g}$&$3.0\times 10^{-21}~\mathrm{m^2}$&\cite{Wee07}\cr
$\sigma_{g0}$&$1.8\sigma_{g}$&\cite{Meirzada17}\cr
$\sigma_{i}$&$(9.5\pm4.7)\times 10^{-21}~\mathrm{m^2}$&\cite{Meirzada17}\cr
$\sigma_{r}$&$(9.8\pm4.9)\times 10^{-21}~\mathrm{m^2}$&\cite{Meirzada17}\cr
$\sigma_{s}$&$(6.1\pm4.4)\times 10^{-23}~\mathrm{m^2}$&\cr
\br
\end{tabular}
\end{indented}
\end{table}

\section{Effective optical depth of the diamond plate}\label{enhancement}

The maximum of transmission of the diamond plate is given by
\begin{equation}
\mathcal{T}_{\mathrm{max}}=\frac{T_de^{-\tau}}{\left(1-R_de^{-\tau}\right)^2},
\end{equation}
where $R_d=\left(\frac{n_d-1}{n_d+1}\right)^2$ and $T_d=1-R_d$ are the   Fresnel coefficients associated to the index of refraction of diamond $n_d$. As $\tau\ll 1$, in the first-order of approximation, we have on one hand $e^{-\tau}\approx1-\tau$, on the other hand $\mathcal{T}_{\mathrm{max}}\approx1-\tau_{\mathrm{eff}}$ where $\tau_{\mathrm{eff}}\ll 1$ corresponds to an effective optical depth taking into account the multiple passes  due to Fresnel reflections within the diamond plate. First-order calculations allow us to write
\begin{equation}
\tau_{\mathrm{eff}}\approx\frac{1+R_d}{1-R_d}\tau,
\end{equation}
which gives $\tau_{\mathrm{eff}}\approx\chi\tau$ with
\begin{equation}
\chi=\frac{n^2_d+1}{2n_d},
\end{equation}
representing the absorption enhancement factor due to Fresnel reflections.

\section{Spectral profile of the optical depth}\label{Odepth}

We assume a Lorentzian shape for the ESR, thus we can write
\begin{equation}
\tau(x)=\tau_{\mathrm{off}}+\frac{\tau_{\rm on}-\tau_{\rm off}}{1+x^2},
\end{equation}
where $x=\frac{2(\nu_{\mathrm{MW}}-\nu_{\rm ESR})}{\Delta\nu_{\mathrm{ESR}}}$ and $\nu_{\rm MW}$ is the frequency of the microwave. We have thus $\frac{\partial\tau}{\partial\nu_{\mathrm{ESR}}}=-\frac{2}{\Delta\nu_{\mathrm{ESR}}}\frac{\partial \tau}{\partial x}$. The maximum of sensitivity is obtained for $x=\frac{1}{\sqrt{3}}$ which gives
\begin{equation}
\left.\frac{\partial \tau}{\partial x}\right|_{x=\frac{1}{\sqrt{3}}}=\frac{3\sqrt{3}(\tau_{\rm off}-\tau_{\rm on})}{8},
\end{equation}
and 
\begin{equation}
\tau\left(\frac{1}{\sqrt{3}}\right)=\frac{\tau_{\mathrm{off}}+3\tau_{\mathrm{on}}}{4}.
\end{equation}
This maximal value of the optical depth is used to determine the optimal value of $\frac{\partial\tau}{\partial\nu_{\rm ESR}}$ given in Eq. (\ref{max_sens}).

\section*{References}
%\bibliography{Biblio_NV}

\end{document}